\documentclass[12pt, centerh1]{article}
\usepackage{natbib}

\usepackage{amsthm,amsmath,natbib}
\RequirePackage[colorlinks,citecolor=blue,urlcolor=blue]{hyperref}

\usepackage{latexsym}
\usepackage{amsmath}
\usepackage{amsfonts}
\usepackage{amssymb}
\usepackage{psfrag}
\usepackage{graphicx}

\usepackage{ctable}
\usepackage{url}
\usepackage{subfigure}
\usepackage{enumitem}
\usepackage{xspace}
\usepackage{xcolor}
\usepackage{color}
\usepackage{colortbl} 
\usepackage{setspace}

\newcommand{\mmtfa}{MMtFA\xspace}
\newcommand{\pgmm}{EPGMM\xspace}
\newcommand{\mcgfa}{MCGFA\xspace}

\newcommand{\R}{\textsf{R}\xspace}

\def\bzero{\boldsymbol{0}}
\def\b1{\boldsymbol{1}}

\def\be{\boldsymbol{e}}

\def\br{\boldsymbol{r}}

\def\bu{\boldsymbol{u}}
\def\bv{\boldsymbol{v}}
\def\bx{\boldsymbol{x}}

\def\bz{\boldsymbol{z}}

\def\bX{\boldsymbol{X}}

\def\bS{\boldsymbol{S}}
\def\bI{\boldsymbol{I}}

\def\bU{\boldsymbol{U}}

\def\bvartheta{\boldsymbol{\vartheta}}

\def\bmu{\boldsymbol{\mu}}
\def\blambda{\boldsymbol{\lambda}}

\def\bbeta{\boldsymbol{\beta}}

\def\btheta{\boldsymbol{\theta}}
\def\bvartheta{\boldsymbol{\vartheta}}

\def\bSigma{\boldsymbol{\Sigma}}

\def\blambda{\boldsymbol{\lambda}}
\def\bLambda{\boldsymbol{\Lambda}}

\def\bPsi{\boldsymbol{\Psi}}
\def\bTheta{\boldsymbol{\Theta}}

\def\diag{\text{diag}}
\def\tr{\text{tr}}

\newcommand{\real}{I\hspace{-0.9mm}R}   

\newcommand{\I}{\boldsymbol{I}}

\renewcommand{\l}{\left}
\renewcommand{\r}{\right}

\newenvironment{equational*}{\equation\aligned}{\endaligned\endequation}

\title{High-dimensional unsupervised classification via parsimonious contaminated mixtures}
\author{Antonio Punzo$^*$, Martin Blostein$^{**}$ and Paul D. McNicholas$^{**}$}
\date{\small $^{*}$Department of Economics and Business, University of Catania, Catania, Italy\\
$^{**}$Department of Mathematics and Statistics, McMaster University, Ontario, Canada.}

\begin{document}
\maketitle
\begin{abstract}
The contaminated Gaussian distribution represents a simple heavy-tailed elliptical generalization of the Gaussian distribution; unlike the often-considered $t$-distribution, it also allows for automatic detection of mild outlying or ``bad'' points in the same way that observations are typically assigned to the groups in the finite mixture model context.
Starting from this distribution, we propose the contaminated factor analysis model as a method for dimensionality reduction and detection of bad points in higher dimensions. 
A mixture of contaminated Gaussian factor analyzers (MCGFA) model follows therefrom, and extends the recently proposed mixture of contaminated Gaussian distributions to high-dimensional data. 
We introduce a family of 32 parsimonious models formed by introducing constraints on the covariance and contamination structures of the general MCGFA model.
We outline a variant of the expectation-maximization algorithm for parameter estimation. 
Various implementation issues are discussed, and the novel family of models is compared to well-established approaches on both simulated and real data.\\[-10pt]

\end{abstract}

\section{Introduction}
\label{sec:Introduction}

Unsupervised classification --- also called cluster analysis or clustering --- is an important subfield of pattern recognition, where the objective is to find homogeneous subpopulations within data \citep{Theo:Kout:Patt:2008}. 
For $p$-dimensional data assumed to arise from a continuous random vector, clustering is commonly focused on elliptical distributions \citep{Camb:Onth:1981} and the Gaussian distribution is the most widely considered elliptical distribution because of its computational and theoretical convenience.  
However, for many practical clustering problems, the tails of the Gaussian distribution are lighter than required to effectively identify homogeneous subpopulations \citep{Nguy:Wu:Zhan:Boun:2014}.
This is often due to the presence of mild outlying or ``bad'' points (see \citealp{Aitk:Wils:Mixt:1980} and \citealp[][pp.~79--80]{Ritt:Robu:2015}), here defined clusterwise \citep{Punz:McNi:Robu:2016} as points that do not really deviate from the Gaussian distribution and are not strongly outlying, but rather they produce an overall within-cluster distribution that is too heavy-tailed to be modeled by the Gaussian \citep{Mazz:Punz:Mixt:2018}. 
		These points are distributed elliptically around the regular clusters 
		and can be dealt with by using heavy-tailed elliptical distributions. 
		Endowed with heavy tails, they offer the flexibility needed for achieving robustness to bad points, whereas
the Gaussian distribution, used as the reference distribution for the typical observations, lacks sufficient fit.
Examples in this direction are the $t$-distribution, thanks to its concentration parameter, i.e., the degrees of freedom (\citealp{Lang:Litt:Tayl:Robu:1989}, \citealp{Kotz:Nada:Mult:2004} and \citealp{Gao:Wen:Wang:Fast:2017}), and the contaminated Gaussian distribution \citep{Tuke:Asur:1960}, a two-component Gaussian mixture in which one of the components, with a large prior probability, represents the ``good'' observations, and the other, with a small prior probability, the same mean, and an inflated covariance matrix, represents the bad observations \citep{Aitk:Wils:Mixt:1980}; in the univariate case, see also \citet{Mazz:Punz:Mode:2019}.  
  
\citet{Punz:McNi:Robu:2016} have recently proposed mixtures of $G$ contaminated Gaussian distributions as a robust generalization of mixtures of Gaussian distributions, and as an alternative to mixtures of $t$ distributions (\citealp{McLa:Peel:Robu:1998}, \citealp{Peel:McLa:Robu:2000}, \citealp{Shoh:Robu:2002}, \citealp{Sfik:Niko:Gala:Robu:2007}, and \citealp{Gao:Wen:Wang:Fast:2017}) and, more in general, to mixtures of elliptical heavy-tailed distributions such as those proposed by \citet{SUN20102447} and \citet{Bagn:Punz:Zoia:Them:2016}.
However, the mixture of $G$ contaminated Gaussian distributions, with unrestricted component-covariance matrices of the good observations, say $\bSigma_1,\ldots,\bSigma_G$, is a highly parametrized model with $p\left(p + 1\right)/2$ parameters for each $\bSigma_g$, $g=1,\ldots,G$. 
To introduce parsimony, \citet{Punz:McNi:Robu:2016} also define fourteen variants of the general model obtained, as in \citet{Cele:Gova:Gaus:1995}, via eigen-decomposition of $\bSigma_1,\ldots,\bSigma_G$. 
This family of models can be fitted in the \textsf{R} software environment for statistical computing and graphics \citep{R:2019} via the \textbf{ContaminatedMixt} package \citep{Punz:Mazz:McNi:Cont:2018}. 
But if $p$ is large relative to the sample size $n$, it may not be possible to use this decomposition to infer an appropriate model for $\bSigma_1,\ldots,\bSigma_G$. 
Even if it is possible, the results may not be reliable due to potential problems with near-singular estimates of $\bSigma_g$ when $p$ is large relative to $n$.

To address this problem, following the literature on the adoption of factor analyzers within mixture models (see, among many others, \citealp[][Chapter~8]{McLa:Peel:fini:2000}, \citealp{McLa:Peel:Bean:Mode:2003}, \citealp{McNi:Murp:Pars:2008}, \citealp{Zhao:Yu:Fast:2008}, \citealp{Mont:Viro:Maxi:2011}, \citealp{WEI20124346}, \citealp{Sube:Punz:Ingr:McNi:Clus:2013,Sube:Punz:Ingr:McNi:Clus:2015}, and \citealp[][Chapter~3]{mcnicholas16a}), we propose mixtures of contaminated Gaussian factor analyzers, where a contaminated Gaussian factor analysis model is used for each mixture component.
The result is a means of fitting mixtures of contaminated Gaussian distributions in situations where $p$ would be sufficiently large (perhaps relative to the sample size $n$) to cause potential problems with singular or near-singular estimates of $\bSigma_1,\ldots,\bSigma_G$.
The number of free parameters is controlled through the dimension of the latent factor space.
Additionally, we propose a family of 32 variants of this model obtained by applying different constraints to the factor loading, error variance matrices (in analogy with \citealp{McNi:Murp:Pars:2008}) and contamination parameters of each mixture component. 
These variants further reduce the number of model parameters, and allow more accurate parameter estimation when mixture components share similar characteristics.

The paper is organized as follows.
Section~\ref{sec:Mixtures of contaminated Gaussian factor analyzers} briefly recalls the contaminated Gaussian distribution (Section~\ref{sec:The contaminated Gaussian distribution}).
It then introduces the contaminated Gaussian factor analysis model (Section~\ref{sec:The contaminated Gaussian factor analysis model}), the mixture of contaminated Gaussian factor analyzers (MCGFA) model, and the family of 32 parsimonious variants of the MCGFA model (Section~\ref{subsec:Parsimonious MCGFA Models}).
This family represents the core of the paper.
Section~\ref{sec:Mixture of contaminated Gaussian factor analyzers: ML estimation via the AECM algorithm} details the alternating expectation-conditional maximization algorithm used for fitting the MCGFA model.
Some computational details are provided in Section~\ref{sec:Computational details}.
In Section~\ref{sec:Comparison with competing methods}, the performance of our family of models is evaluated with respect to two alternative parsimonious family of models through several simulated and real data analyses.
Computationally, the heavy lifting is done in the \textsf{C} programming language, with an \textsf{R} interface, and an {\sf R} package will shortly be released.
The paper concludes with a discussion in Section~\ref{sec:Discussion and future work}.

\section{Mixtures of Contaminated Gaussian Factor Analyzers}
\label{sec:Mixtures of contaminated Gaussian factor analyzers}

\subsection{The contaminated Gaussian distribution}
\label{sec:The contaminated Gaussian distribution}

The $p$-variate random vector $\bX$ is said to have a contaminated Gaussian distribution \citep{Tuke:Asur:1960} with mean $\bmu$, scale matrix $\bSigma$, proportion of good points $\alpha\in\left(0,1\right)$, and degree of contamination $\eta>1$, if its probability density function (pdf) is given by
\begin{equation}
p_{\text{CN}}\left(\bx;\bmu,\bSigma,\alpha,\eta\right)=\alpha p_{\text{N}}\left(\bx;\bmu,\bSigma\right)+\left(1-\alpha\right) p_{\text{N}}\left(\bx;\bmu,\eta\bSigma\right),
\label{eq:contaminated Gaussian distribution}
\end{equation}
where $p_{\text{N}}\left(\cdot;\bmu,\bSigma\right)$ denotes the pdf of a $p$-variate normal distribution with mean $\bmu$ and covariance matrix $\bSigma$.
If $\bX$ has the pdf in \eqref{eq:contaminated Gaussian distribution}, then we write $\bX\sim \mathcal{CN}_{p}\left(\bmu,\bSigma,\alpha,\eta\right)$.
As we can see in \eqref{eq:contaminated Gaussian distribution}, a contaminated Gaussian distribution  is a two-component Gaussian mixture in which one of the components, typically with a large prior probability $\alpha$, represents the ``good'' observations, and the other, with a small prior probability, the same mean, and an inflated covariance matrix $\eta\bSigma$, represents the ``bad'' observations \citep{Aitk:Wils:Mixt:1980}.
As a special case of \eqref{eq:contaminated Gaussian distribution}, if $\alpha$ and $\eta$ tend to one, we obtain the Gaussian distribution with mean $\bmu$ and covariance matrix $\bSigma$,
i.e., $\bX\sim \mathcal{N}_{p}\left(\bmu,\bSigma\right)$.

As for the popular $t$ distribution, the contaminated Gaussian distribution can be also seen as a special case of the Gaussian scale mixture
\begin{equation}
\int_{S_h} p_{\text{N}}\left(\bx;\bmu,\bSigma/w\right)h\left(w;\btheta\right)dw,
\label{eq:Gaussian scale mixture model}
\end{equation}
where $h\left(w;\btheta\right)$ is the mixing probability density (or mass) function, with support $S_h \subseteq \real_{>0}$, depending on the parameter(s) $\btheta$.
The pdf in \eqref{eq:Gaussian scale mixture model} is unimodal, elliptically symmetric, and heavier tailed than the Gaussian distribution (see, e.g., \citealp{Barn:Kent.Sore:Inte:Norm:1982}, \citealp{Wata:Yama:TheE:2003} \citealp[][Section~2.6]{Fang:Kotz:Ng:Symm:2013}, \citealp{Yama:Robu:2004} and \citealp[][Section~7.4]{McLa:Peel:fini:2000}).
The tail weight of the Gaussian scale mixture distribution is governed by $\btheta$.
In detail, the contaminated Gaussian distribution is a special case of \eqref{eq:Gaussian scale mixture model} if we consider the dichotomous random variable
\begin{equation}
W=\left\{
\begin{array}{ll}
1 & \text{with probability $\alpha$},\\
1/\eta	& \text{with probability $1-\alpha$},
\end{array}\right. 
\label{eq:W}
\end{equation}
with probability mass function
\begin{equation}
h\left(w;\btheta\right)=\alpha^{\frac{w-1/\eta}{1-1/\eta}}\left(1-\alpha\right)^{\frac{1-w}{1-1/\eta}},
\label{eq:C}
\end{equation}
where $\btheta=\left(\alpha,\eta\right)$.
Advantageously, the Gaussian scale mixture representation of $\bX\sim \mathcal{CN}_{p}\left(\bmu,\bSigma,\alpha,\eta\right)$ can be expressed hierarchically as
\begin{align}
W &\sim \mathcal{C}\left(\alpha,\eta\right),
\label{eq:W C}
\\
\bX|w &\sim \mathcal{N}_{p}\left(\bmu,\bSigma/w\right),
\label{eq:X|w N}
\end{align}
where $\mathcal{C}\left(\alpha,\eta\right)$ denotes the dichotomous contamination variable defined by \eqref{eq:W} and \eqref{eq:C}.

An advantage of model~\eqref{eq:contaminated Gaussian distribution} with respect to the existing Gaussian scale mixtures is that, once the parameters in $\boldsymbol{\vartheta}=\left\{\bmu,\bSigma,\alpha,\eta\right\}$ are estimated, say $\hat{\boldsymbol{\vartheta}}=\{\hat{\boldsymbol{\mu}},\hat{\bSigma},\hat{\alpha},\hat{\eta}\}$, we can establish whether a generic point $\bx$ is either good or bad via its \textit{a~posteriori} probability. 
That is, compute
\begin{equation}
P(\text{$\bx$ is good}|\hat{\boldsymbol{\vartheta}})=\frac{\hat{\alpha} p_{\text{N}}(\bx;\hat{\bmu},\hat{\bSigma})}{p_{\text{CN}}(\bx;\hat{\boldsymbol{\vartheta}})},
\label{eq:probability good}
\end{equation}
and consider $\bx$ as good if $P(\text{$\bx$ is good}|\hat{\boldsymbol{\vartheta}})>1/2$.

\subsection{The contaminated Gaussian factor analysis model}
\label{sec:The contaminated Gaussian factor analysis model}

The (Gaussian) factor analysis model \citep{Spea:Thep:1904,bartlett53,lawley62,lawley71} is a well-known, and widely used, data reduction tool aiming to find latent factors that explain the variability in the data. 
Suppose we have $\bX_1,\ldots,\bX_n$ from a factor analysis model. The model \citep[see][Chapter~3]{Bart:Knot:Mous:Late:2011} assumes that the $p$-variate random vector $\bX_i$ is modelled using a $q$-variate vector of factors $\bU_i\sim \mathcal{N}_q\left(\boldsymbol{0}_q,\bI_q\right)$, where $q< p$ and the $\bU_i$ are independently distributed.
The model is
\begin{equation}
\bX_i = \bmu + \bLambda \bU_i + \be_i, 
\label{eq:Gaussian factor model}
\end{equation}
where $\bLambda$ is a $p\times q$ matrix of factor loadings, $\be_i\sim \mathcal{N}_p\left(\boldsymbol{0}_p,\boldsymbol{\Psi}\right)$ is the error term, with $\bPsi=\diag\left(\psi_1,\ldots,\psi_p\right)$, and the $\be_i$ are independently distributed and independent of the $\bU_i$.
It follows from \eqref{eq:Gaussian factor model} that $\bX_i\sim \mathcal{N}_p\left(\bmu,\bLambda\bLambda'+\bPsi\right)$.

The factor analysis model is, however, sensitive to bad points as it adopts the Gaussian distribution for errors and latent factors. 
To improve its robustness, for data having longer than Gaussian tails or bad points, \citet{McLa:Bean:BenT:Exte:2007} introduce the $t$-factor analysis model which considers the multivariate $t$ for the distributions of the errors and the latent factors \citep[see also][]{andrews11a}.
We extend this branch of literature by introducing the contaminated Gaussian factor analysis model.  

Based on \eqref{eq:Gaussian factor model}, the contaminated Gaussian factor analysis model generalizes the corresponding Gaussian factor analysis model by assuming
\begin{equation}
\begin{pmatrix} 
\bX_i \\ 
\bU_i  
\end{pmatrix}\sim \mathcal{CN}_{p+q}\left(\bmu^*,\bSigma^*,\alpha,\eta\right),
\label{eq:Factor: joint density of X and U}
\end{equation}
where 
\begin{equation*}
\bmu^*=
\begin{pmatrix} 
\bmu \\ 
\bzero_q  
\end{pmatrix} \quad \text{and} \quad 
\bSigma^*
= 
\begin{pmatrix}  
\bLambda\bLambda'+\bPsi & \bLambda\\
 \bLambda' & \bI_q \\
\end{pmatrix}. 
\end{equation*}
Using the Gaussian scale mixture \textcolor{blue}{representation} of the contaminated Gaussian distribution discussed in Section~\ref{sec:The contaminated Gaussian distribution}, the joint pdf of $\bX_i$ and $\bU_i$, given $W_i=w_i$, can be written
\begin{equation}
\begin{pmatrix} 
\bX_i \\ 
\bU_i 
\end{pmatrix}
\Bigg|w_i \sim \mathcal{N}_{p+q}\left(\bmu^*,\bSigma^*/w_i\right),
\label{eq:Factor: joint density of X and U given w}
\end{equation}
with $W_i \sim \mathcal{C}\left(\alpha,\eta\right)$.
Thus,
\begin{align*}
\bX_i|w_i   & \sim   \mathcal{N}_p\left(\bmu,\left(\bLambda\bLambda'+\bPsi\right)/w_i\right),\\
\bU_i|w_i   & \sim   \mathcal{N}_q\left(\bzero_q,\bI_q/w_i\right),\\
\be_i|w_i   & \sim   \mathcal{N}_p\left(\bzero_p,\bPsi/w_i\right),
\end{align*}
so that
\begin{align*}
\bX_i  &\sim  \mathcal{CN}_p\left(\bmu,\bLambda\bLambda'+\bPsi,\alpha,\eta\right),\\
\bU_i  &\sim  \mathcal{CN}_q\left(\bzero_q,\bI_q,\alpha,\eta\right),\\
\be_i  &\sim  \mathcal{CN}_p\left(\bzero_p,\bPsi,\alpha,\eta\right).
\end{align*}
The factors $\bU_i$ and error terms $\be_i$ are no longer independently distributed as in the usual Gaussian factor analysis model; however, they remain uncorrelated.

\subsection{Parsimonious MCGFA models}
\label{subsec:Parsimonious MCGFA Models}

To robustify the classical mixture of Gaussian distributions to the occurrence of bad points, and also to allow for their automatic detection (see \citealp{Zime:Schu:Krie:Asur:2012}, \citealp{Pime:Clif:Clif:Tara:Arev:2014}, \citealp{Domi:Fili:Mich:Zoua:Acom:2018} for recent surveys about outlier detection methods), \citet{Punz:McNi:Robu:2016} propose the mixture of contaminated Gaussian distributions
\begin{equation}
p\left(\bx;\boldsymbol{\vartheta}\right)=\sum_{g=1}^G\pi_gp_{\text{CN}}(\bx;\bmu_g,\bSigma_g,\alpha_g,\eta_g)
\label{eq:mixture of multivariate contaminated Gaussian distributions}
\end{equation}
where, for the $g$th mixture component, $\pi_g>0$ is its mixing proportion, with $\sum_{g=1}^G\pi_g=1$, and the density $p_{\text{CN}}(\bx;\bmu_g,\bSigma_g,\alpha_g,\eta_g)$ is defined as in \eqref{eq:contaminated Gaussian distribution}.
For recent extensions of model~\eqref{eq:mixture of multivariate contaminated Gaussian distributions} to the hidden Markov model and regression setting, see \citet{Punz:Maru:Clus:2016}, \citet{Maru:Punz:Mode:2016}, \citet{Punz:McNi:Robu:2017} and \citet{Mazz:Punz:Mixt:2018}.

In \eqref{eq:mixture of multivariate contaminated Gaussian distributions}, there are $p\left(p + 1\right)/2$ parameters for each $\boldsymbol{\Sigma}_g$, $g=1,\ldots,G$. 
This means that, as the number of components $G$ grows, the total number of free parameters can quickly become very large leading to overfitting.
To model high-dimensional data, and to add parsimony, we consider the contaminated Gaussian factor analysis model of Section~\ref{sec:The contaminated Gaussian factor analysis model} in each mixture component; this leads to the mixture of contaminated Gaussian factor analyzers given by \eqref{eq:mixture of multivariate contaminated Gaussian distributions} but with the component scale matrices given by
\begin{equation}
\boldsymbol{\Sigma}_g=\bLambda_g\bLambda_g'+\bPsi_g.
\label{eq:restriction}
\end{equation}

Following the work of \citet{McNi:Murp:Pars:2008} on mixtures of Gaussian factor analyzers, and of \citet{andrews11a,andrews11c}, \citet{steane12} and \citet{lin14} on mixtures of $t$ factor analyzers, we introduce a unified family of 32 mixtures of contaminated Gaussian factor analyzers by imposing five different sets of constraints, three on the covariance structure parameters $\{\bLambda_g\}_{g=1}^G$ and $\{\bPsi_g\}_{g=1}^G$, and the remaining two on the contamination parameters $\{\alpha_g\}_{g=1}^G$ and $\{\eta_g\}_{g=1}^G$.  
First, the factor loading matrices $\bLambda_g$ may be constrained to be equal across groups, i.e., $\bLambda_g=\bLambda$; this situation is sometimes referred to as $\bLambda_g$ being ``tied'' but we shall use the term ``constrained'' herein.
This constraint prevents local dimensionality reduction, but if the mixture components indeed share similar covariance structures, provides a simpler model and greater stability for parameter estimation. 
Second, the error variance matrices $\bPsi_g$ may be constrained across groups; this is consistent with the interpretation of $\bPsi$ as sensor noise that affects all observations in the same way \citep[see][]{ghahramani97}. 
Third, we may assume that error variances in each variable are the same within each group, or that we have \textit{isotropic} errors \citep[see][]{tipping99b}.
Finally, we may set equal across groups either the proportions of good observations $\alpha_g$ or the inflation parameters $\eta_g$.
So all together, the possible constraints are:
\begin{enumerate}
	\item loading matrices constrained across groups, i.e., $\bLambda_1 = \cdots = \bLambda_G = \bLambda$;
	\item error variance matrices constrained across groups, i.e., $\bPsi_1 = \cdots = \bPsi_G = \bPsi$;
	\item isotropic errors within groups, i.e., $\bPsi_g = \psi_g\I_p$, $\psi_g\in\mathbb{R}^+$;
	\item proportions of good observations constrained across groups, i.e., $\alpha_1 = \cdots = \alpha_G = \alpha$;
	\item inflation parameters constrained across groups, i.e., $\eta_1 = \cdots = \eta_G = \eta$.
\end{enumerate}
Each constraint may be applied or not, independently of the others, yielding 32 models. 
The models are for simplicity labeled by merging two groups of letter codes: the first group having three letters referring to the constraints on the covariance structure, and the second group with two letters referring to the constraints on the contamination parameters.
Each letter can be C and U, where U indicates unconstrained and C indicates constrained.   
Thus the unconstrained, or most general, MCGFA model is denoted UUUUU. 
The full MCGFA family of models is presented in \tablename~\ref{tab:Family of models}, along with their number of free parameters, denoted $\#\text{par}_\text{cov}$ and $\#\text{par}_\text{cont}$, related to the scale matrices $\bSigma_1,\ldots,\bSigma_G$, and to contamination parameters $\alpha_1,\ldots,\alpha_G$ and $\eta_1,\ldots,\eta_G$, respectively.
Note that the overall number of free parameters, denoted $\#\text{par}$, in any of the 32 model variants is $\left(G-1\right)+ Gp + \#\text{par}_\text{cov}+\#\text{par}_\text{cont}$.   
\begin{table}[!ht]
	\centering
	\singlespacing
	\caption{Nomenclature and parsimonious structures for members of the MCGFA family, where
	the number of free covariance parameters is denoted by $\#\text{par}_\text{cov}$ and the number of contamination parameters is denoted by $\#\text{par}_\text{cont}$.}
	\label{tab:Family of models}
	\begin{tabular}{ccc cc cc}
		\toprule
		$\bLambda_g = \bLambda$ & $\bPsi_g = \bPsi$ & $\bPsi_g = \psi_g\I$ & $\alpha_g = \alpha$ & $\eta_g = \eta$ & $\#\text{par}_\text{cov}$ & $\#\text{par}_\text{cont}$\\
		\midrule
		C & C & C & C & C & $pq-q(q-1)/2+1$ & $2$ \\[1.5mm]
		
		C & C & C & C & U & $pq-q(q-1)/2+1$ & $G+1$ \\
		C & C & C & U & C & $pq-q(q-1)/2+1$ & $G+1$ \\
		C & C & U & C & C & $pq-q(q-1)/2+p$ & $2$  \\
		C & U & C & C & C & $pq-q(q-1)/2+G$ & $2$  \\ 
		U & C & C & C & C & $G[pq-q(q-1)/2]+1$ & $2$  \\[1.5mm]
		
		C & C & C & U & U & $pq-q(q-1)/2+1$ & $2G$ \\
		C & C & U & C & U & $pq-q(q-1)/2+p$ & $G+1$  \\
		C & U & C & C & U & $pq-q(q-1)/2+G$ & $G+1$  \\ 
		U & C & C & C & U & $G[pq-q(q-1)/2]+1$ & $G+1$  \\
		C & C & U & U & C & $pq-q(q-1)/2+p$ & $G+1$  \\
		C & U & C & U & C & $pq-q(q-1)/2+G$ & $G+1$  \\ 
		U & C & C & U & C & $G[pq-q(q-1)/2]+1$ & $G+1$  \\
		C & U & U & C & C & $pq-q(q-1)/2+Gp$ & $2$  \\
		U & C & U & C & C & $G[pq-q(q-1)/2]+p$ & $2$  \\
		U & U & C & C & C & $G[pq-q(q-1)/2]+G$ & $2$  \\[1.5mm]
		
		C & C & U & U & U & $pq-q(q-1)/2+p$ & $2G$  \\
		C & U & C & U & U & $pq-q(q-1)/2+G$ & $2G$  \\ 
		C & U & U & C & U & $pq-q(q-1)/2+Gp$ & $G+1$  \\
		C & U & U & U & C & $pq-q(q-1)/2+Gp$ & $G+1$  \\
		U & C & C & U & U & $G[pq-q(q-1)/2]+1$ & $2G$  \\
		U & C & U & C & U & $G[pq-q(q-1)/2]+p$ & $G+1$  \\
		U & C & U & U & C & $G[pq-q(q-1)/2]+p$ & $G+1$  \\
		U & U & C & C & U & $G[pq-q(q-1)/2]+G$ & $G+1$  \\
		U & U & C & U & C & $G[pq-q(q-1)/2]+G$ & $G+1$  \\
		U & U & U & C & C & $G[pq-q(q-1)/2]+Gp$ & $2$ \\[1.5mm]
		
		C & U & U & U & U & $pq-q(q-1)/2+Gp$ & $2G$  \\
		U & C & U & U & U & $G[pq-q(q-1)/2]+p$ & $2G$  \\
		U & U & C & U & U & $G[pq-q(q-1)/2]+G$ & $2G$  \\
		U & U & U & C & U & $G[pq-q(q-1)/2]+Gp$ & $G+1$ \\
		U & U & U & U & C & $G[pq-q(q-1)/2]+Gp$ & $G+1$ \\[1.5mm]
		
		U & U & U & U & U & $G[pq-q(q-1)/2]+Gp$ & $2G$ \\
		\bottomrule
	\end{tabular}
\end{table}

\subsection{Model selection}
\label{subsec:Model selection}

As usual in the literature about mixture models \citep{fraley98}, we handle model order selection (estimating the number of mixture components $G$), factorial dimension selection (determining the number of latent factors $q$), and model structure selection (determining the best parsimonious structure among those in \tablename~\ref{tab:Family of models}), simultaneously by the Bayesian information criterion \citep[BIC;][]{Schw:Esti:1978}: 
\begin{equation}
\text{BIC}=-2l(\hat{\boldsymbol{\vartheta}})+\#\text{par}\times\ln n,
\label{eq:BIC}
\end{equation}
where $l(\hat{\boldsymbol{\vartheta}})$ is the maximized (observed-data) log-likelihood and $n$ is the sample size; for more recent alternatives to the BIC see, e.g., \citet{Mehr:Hoss:Araa:Impr:2016}.
Note that, when formulated as in \eqref{eq:BIC}, models with smaller BIC values are preferred.
\citet{Lero:Cons:1992} and \citet{Roed:Wass:Prac:1997} established the consistency of the BIC for mixture models.

%
However, when the number of variables $p$ is very large, the BIC may grossly underestimate the order $G$ \citep[see, e.g.,][]{bhattacharya14}.
As well-documented in \citet{Grah:Mill:Unsu:2006}, this failure is not mainly attributable to the criterion, but rather to the lack of ``structure''.
This problem roughly amounts to inadequate number of members, for each fixed value of $G$, in the considered family of models. 
We try to mitigate this problem by searching over a rich family of 32 parsimonious models and by applying dimensionality reduction simultaneously to clustering.

%

\section{Maximum likelihood estimation via the AECM algorithm}
\label{sec:Mixture of contaminated Gaussian factor analyzers: ML estimation via the AECM algorithm}

To find ML estimates for the parameters $\boldsymbol{\vartheta}=\left\{\pi_g,\bmu_g,\bLambda_g,\bPsi_g,\alpha_g,\eta_g\right\}_{g=1}^G$ of the MCGFA model, we consider the application of the alternating expectation-conditional maximizations (AECM) algorithm of \citet{meng97}.
The AECM algorithm is an extension of the expectation-conditional maximization (ECM) algorithm \citep{meng93}, where the specification of the complete data is allowed to be different on each CM-step. 
The ECM algorithm is itself a variant of the classical expectation-maximization (EM) algorithm \citep{dempster77}, which is a natural approach for ML estimation when there are sources of latent or hidden data.
In our case, we have two sources of latent data: the component membership of each observation, and the classification of each observation as good or bad within each component.
To denote the first source, we use $\bz_1,\ldots,\bz_n$, where $\bz_i=\left(z_{i1},\ldots,z_{iG}\right)'$ so that $z_{ig}=1$ if observation~$i$ is in component $g$, and $z_{ig}=0$ otherwise.
For the second source, we use the indicator variable 
$$
V=\frac{W-1/\eta}{1-1/\eta},
$$ 
which is a linear transformation of $W$ in \eqref{eq:W}.
This yields $\bv_1,\ldots,\bv_n$, where $\bv_i=\left(v_{i1},\ldots,v_{iG}\right)'$ so that $v_{ig}=1$ if observation~$i$ in group~$g$ is good and $v_{ig}=0$ if observation $i$ in group $g$ is bad.

To apply the AECM algorithm, we partition 
$\bvartheta=\left\{\bvartheta_1,\bvartheta_2\right\}$, where $\bvartheta_1=\left\{\pi_g,\bmu_g,\alpha_g,\eta_g\right\}_{g=1}^G$ and $\boldsymbol{\vartheta}_2=\left\{\bLambda_g,\bPsi_g\right\}_{g=1}^G$, so that the complete-data likelihood is easy to maximize for $\bvartheta_1$ given $\bvartheta_2$ and \textit{vice~versa}.
Therefore, the $\left(k+1\right)$th iteration of our AECM algorithm consists of two cycles: there is one E-step and \textcolor{orange}{two CM-steps} for the first cycle and one E-step and one CM-step for the second cycle. 
The two cycles correspond to the partition of $\bvartheta$ into $\bvartheta_1$ and $\bvartheta_2$.
The two CM-steps of the first cycle correspond to the partition of $\bvartheta_1$ as $\bvartheta_1=\left\{\bvartheta_{11},\bvartheta_{12}\right\}$, where $\bvartheta_{11}=\left\{\pi_g,\bmu_g,\alpha_g\right\}_{g=1}^G$ and $\boldsymbol{\vartheta}_{12}=\left\{\eta_g\right\}_{g=1}^G$.

All maximization steps in the algorithm are solvable analytically. 
Thus all parameter updates are available in closed form, avoiding any use of numerical optimization. This stands in contrast to the lack of a closed form update for the degrees of freedom in the case of the $t$ distribution.

\subsection{First cycle}
\label{subsec:First cycle}

For the first cycle of the AECM algorithm, we specify the missing data to be $\bz_1,\ldots,\bz_n$ and $\bv_1,\ldots,\bv_n$.
Thus, the complete data are $\left(\bx'_1,\ldots,\bx'_n,\bz'_1,\ldots,\bz'_n,\bv'_1,\ldots,\bv'_n\right)$ and the complete-data log-likelihood can be written as 
\begin{equation*}
l_{1c}(\bvartheta_1) =
l_{1c_1}(\{\pi_g\}_{g=1}^G) 
+ 
l_{1c_2}(\{\alpha_g\}_{g=1}^G)
+
l_{1c_3}(\{\bmu_g,\eta_g\}_{g=1}^G),
\end{equation*}
where
\begin{align}
	l_{1c_1}(\{\pi_g\}_{g=1}^G) = & \sum_{i=1}^n\sum_{g=1}^Gz_{ig}\log\pi_g\nonumber\\
	l_{1c_2}(\{\alpha_g\}_{g=1}^G) = & \sum_{i=1}^n\sum_{g=1}^Gz_{ig}\left[v_{ig}\log\alpha_g+(1-v_{ig})\log(1-\alpha_g)\right]\nonumber\\
	l_{1c_3}(\{\bmu_g,\eta_g\}_{g=1}^G) = & -\frac{1}{2} \sum_{i=1}^n\sum_{g=1}^G\bigg[z_{ig}\log|\boldsymbol{\Sigma}_g^{(k)}|+p z_{ig}(1-v_{ig})\log\eta_g\nonumber\\
	& \qquad\quad +z_{ig}\left(v_{ig}+\frac{1-v_{ig}}{\eta_g}\right)(\bx_i-\bmu_g)' (\bSigma_g^{(k)})^{-1}(\bx_i-\bmu_g)\bigg],
	\label{eq:lc3}
\end{align}
where $\bSigma_g^{(k)}=\bLambda_g^{(k)}\bLambda_g^{(k)'}+\bPsi_g^{(k)}$.
In \eqref{eq:lc3}, constants with respect to the parameters are omitted for the sake of brevity.

\subsubsection{E-step}
\label{subsubsec:E-step 1}

The E-step on the first cycle of the $\left(k + 1\right)$th iteration requires the calculation of the expectation of $l_{1c}$ given the observed data $\bx_1,\ldots,\bx_n$ and $\bvartheta^{(k)}$.
To do this, we replace $z_{ig}$ with
\begin{equation*}
z_{ig}^{(k )} =
E[Z_{ig}\mid \bx_i,\bvartheta^{(k)}]=
\frac{\pi_g^{(k)} p_{\text{CN}}(\bx_i;\bmu_g^{(k )},\bSigma_g^{(k )},\alpha_g^{(k )},\eta_g^{(k )})}{\displaystyle\sum_{j=1}^G\pi_j^{(k)} p_{\text{CN}}(\bx_i;\bmu_j^{(k )},\bSigma_j^{(k )},\alpha_j^{(k )},\eta_j^{(k )})},
\end{equation*}
 and $v_{ig}$ with 
\begin{equation*}
v_{ig}^{(k)} =
E[V_{ig}\mid Z_{ig}=1,\bx_i,\bvartheta^{(k)}] =
\frac{\alpha_g^{(k)} p_{\text{N}}(\bx_i; \bmu_g^{(k)},\bSigma_g^{(k)})}{p_{\text{CN}}(\bx_i;\bmu_g^{(k)},\bSigma_g^{(k)},\alpha_g^{(k)},\eta_g^{(k)})},
\end{equation*}
where $Z_{ig}$ and $V_{ig}$ are the random variables related to $z_{ig}$ and $v_{ig}$, respectively.

\subsubsection{CM-step 1}

At the first CM-step on the first cycle of the $(k+1)$th iteration, we maximize the expectation of the complete-data log-likelihood with respect to $\bvartheta_{11}$, fixing $\bvartheta_{12}=\bvartheta_{12}^{(k)}$.
Some algebra yields the following updates for $\pi_g$ and $\bmu_g$:
\begin{eqnarray}
\pi_g^{(k+1)} &=& n_g^{(k)}/n,\nonumber\\
\bmu_g^{(k+1)} &=& \frac{\displaystyle\sum_{i=1}^nz_{ig}^{(k)}\left(v_{ig}^{(k)}+\frac{1-v_{ig}^{(k)}}{\eta_g^{(k)}}\right)\bx_i}{\displaystyle\sum_{i=1}^nz_{ig}^{(k)}\left(v_{ig}^{(k)}+\frac{1-v_{ig}^{(k)}}{\eta_g^{(k)}}\right)},
\label{eq:mu}
\end{eqnarray}
where $n_g^{(k)}=\displaystyle\sum_{i=1}^nz_{ig}^{(k)}$.

As concerns the update of the proportion of good observations $\alpha_g$, we have to distinguish the unconstrained case and the case of tied proportions across groups.
Moreover, for the sake of interpretation, we could require that these proportions should lie within the interval $(\alpha^*,1)$, where $\alpha^*$ is the minimum proportion of good observations.
For the analyses herein we use $\alpha^*=0.5$; this choice is justified by the fact that robust (clustering) techniques typically allow for a contamination rate of at most 50\% \citep{Garc:Gord:Matr:Mayo:Agen:2008,Ritt:Robu:2015}. 
The motivation lies in the (sometimes implicit) assumption that the ``good'' population should correspond to the majority of data.
According to these considerations, in the unconstrained case the update for $\alpha_g$ is
\begin{equation*}
\alpha_g^{(k+1)}=\max\left\{\alpha^*,\frac{1}{n_g^{(k)}}\displaystyle\sum_{i=1}^nz_{ig}^{(k)}v_{ig}^{(k)}\right\},
\end{equation*}
while in the constrained case the update for the common proportion $\alpha$ is
\begin{equation*}
\alpha^{(k+1)}=\max\left\{\alpha^*,\frac{1}{n}\displaystyle\sum_{g=1}^G\sum_{i=1}^nz_{ig}^{(k)}v_{ig}^{(k)}\right\}.
\end{equation*}

\subsubsection{CM-step 2}

At the second CM-step on the first cycle of the $(k+1)$th iteration, we maximize the expectation of the complete-data log-likelihood with respect to $\eta_g$ or $\eta$, depending on the model being fitted, fixing $\bvartheta_{11}=\bvartheta_{11}^{(k+1)}$.
In the less parsimonious ``$\eta_g$'' case, this yields the update
\begin{equation}
\eta_g^{(k+1)} = \max\l\{ \eta^*, \frac{b_g^{(k)}}{p a_g^{(k)}} \r\},
\end{equation}
where
\begin{align*}
a_g^{(k)} &= \sum_{i = 1}^n z_{ig}^{(k)}(1 - v_{ig}^{(k)}),\\
b_g^{(k)} &= \sum_{i=1}^n z_{ig}^{(k)} (1 - v_{ig}^{(k)})(\bx_i-\bmu_g^{(k+1)})'(\bSigma_g^{(k)})^{-1}(\bx_i-\bmu_g^{(k+1)}), 
\end{align*}
and $\eta^*$ is a number close to 1 from the right; for the analyses herein, we use $\eta^*=1.001$.
In the more parsimonious ``$\eta$'' case, the update becomes
\begin{equation}
\eta^{(k+ 1)} = \max\l\{ \eta^*, \frac{b^{(k)}}{p a^{(k)}} \r\},
\end{equation}
where $a^{(k)}=\displaystyle\sum_{g=1}^Ga_g^{(k)}$ and $b^{(k)}=\displaystyle\sum_{g=1}^Gb_g^{(k)}$.

\subsection{Second cycle}
\label{subsec:Second cycle}

For the second cycle of the AECM algorithm, we specify the missing data to be $\bz_1,\ldots,\bz_n$, $\bv_1,\ldots,\bv_n$, and the latent factors $\bu_1,\ldots,\bu_n$.
Therefore, the complete-data log-likelihood can be written as 
\begin{align}
\hspace{-8mm}
l_{2c}(\bvartheta_2) = & 
C + \sum_{g=1}^G\Biggl\{-\frac{n_g}{2}\log|\bPsi_g|-\frac{n_g}{2}\tr(\bPsi_g^{-1}\bS_g^{(k+1)})+\sum_{i=1}^n z_{ig}\left(v_{ig}+\frac{1-v_{ig}}{\eta_g^{(k+1)}}\right)(\bx_i-\bmu_g^{(k+1)})'\bPsi_g^{-1}\bLambda_g\bu_{ig}\nonumber\\
& - \frac{1}{2} \tr\left[\bLambda_g'\bPsi_g^{-1}\bLambda_g\sum_{i=1}^n z_{ig}\left(v_{ig}+\frac{1-v_{ig}}{\eta_g^{(k+1)}}\right)\bu_{ig}\bu_{ig}'\right]\Biggl\},
\label{eq:complete-data loglik 2}
\end{align}
where $n_g=\displaystyle\sum_{i=1}^nz_{ig}$, $C$ is a constant with respect to $\bvartheta_2$, and
\begin{equation}
\bS_g^{(k+1)} =  \frac{1}{n_g}\displaystyle\sum_{i=1}^n z_{ig}\left(v_{ig}+\frac{1-v_{ig}}{\eta_g^{(k+1)}}\right)(\bx_i-\bmu_g^{(k+1)})(\bx_i-\bmu_g^{(k+1)})'.
\label{eq:weighted S}
\end{equation}  

\subsubsection{E-step}

The E-step on the second cycle of the $(k + 1)$th iteration requires the calculation of the expectation of $l_{2c}$ given the observed data and $\bvartheta^{(k+1/2)}=\{\bvartheta_1^{(k+1)},\bvartheta_2^{(k)}\}$.
Operationally, this involves the substitution of $z_{ig}$ and $v_{ig}$ in \eqref{eq:complete-data loglik 2} and \eqref{eq:weighted S} with $z_{ig}^{(k+1/2)}$ and $v_{ig}^{(k+1/2)}$, respectively; the notation changes, with respect to $z_{ig}^{(k)}$ and $v_{ig}^{(k)}$ in Section~\ref{subsubsec:E-step 1}, because we now use the updates $\pi_g^{(k+1)}$, $\alpha_g^{(k+1)}$, $\bmu_g^{(k+1)}$, and $\eta_g^{(k+1)}$ from the first cycle of the algorithm.
The E-step also involves the computation of the following conditional expectations 
\begin{align*}
	& 
E_{\bvartheta^{(k+1/2)}}\left[Z_{ig}\left(V_{ig}+\frac{1-V_{ig}}{\eta_g^{(k+1)}}\right)\bU_{ig}~\Big|~\bx_i\right] = z_{ig}^{(k+1/2)}\left(v_{ig}^{(k+1/2)}+\frac{1-v_{ig}^{(k+1/2)}}{\eta_g^{(k+1)}}\right)\bbeta_g^{(k)}(\bx_i-\bmu_g^{(k+1)}),\\
	& 
	E_{\bvartheta^{(k+1/2)}}[Z_{ig}V_{ig}\bU_{ig}\bU_{ig}'\mid \bx_i] = z_{ig}^{(k+1/2)}v_{ig}^{(k+1/2)}\left[\bI_q-\bbeta_g^{(k)}\bLambda_g^{(k)}+\bbeta_g^{(k)}(\bx_i-\bmu_g^{(k+1)})(\bx_i-\bmu_g^{(k+1)})'\bbeta_g^{(k)'}\right],\\
	& 
	E_{\bvartheta^{(k+1/2)}}\left[Z_{ig}\left(\frac{1-V_{ig}}{\eta_g^{(k+1)}}\right)\bU_{ig}\bU_{ig}'~\Big|~\bx_i\right]\\ &\qquad\qquad\qquad = z_{ig}^{(k+1/2)}\left(\frac{1-v_{ig}^{(k+1/2)}}{\eta_g^{(k+1)}}\right)\left[\bI_q-\bbeta_g^{(k)}\bLambda_g^{(k)} +\bbeta_g^{(k)}(\bx_i-\bmu_g^{(k+1)})(\bx_i-\bmu_g^{(k+1)})'\bbeta_g^{(k)'}\right],
\end{align*} 
where $\bbeta_g^{(k)}=\bLambda_g^{(k)'}(\bLambda_g^{(k)}\bLambda_g^{(k)'}+\bPsi_g^{(k)})^{-1}$.
The precise formula for $\bbeta_g$ changes depending on which constraints are imposed upon $\{\bLambda_g\}_{g=1}^G$ and $\{\bPsi_g\}_{g=1}^G$.
The formulae for each of the eight parsimonious models regarding the covariance structure can be found in \citet[][Appendix~A]{McNi:Murp:Pars:2008}.
It follows that the expected complete-data log-likelihood, omitting the constant terms, is
\begin{equation}\begin{split}
Q_2(\bvartheta_2) =& 
\sum_{g=1}^Gn_g^{(k+1/2)}\bigg\{\frac{1}{2}\log|\bPsi_g^{-1}|\\& -\frac{1}{2}\tr\left(\bPsi_g^{-1}\bS_g^{(k+1)}\right) + \tr\left(\bPsi_g^{-1}\bLambda_g\bbeta_g^{(k)}\bS_g^{(k+1)}\right)
- \frac{1}{2} \tr\left(\bLambda_g'\bPsi_g^{-1}\bLambda_g\bTheta_g^{(k+1/2)}\right)\bigg\},
\label{eq:Q}
\end{split}
\end{equation}
where $n_g^{(k+1/2)}=\displaystyle\sum_{i=1}^nz_{ig}^{(k+1/2)}$ and $\bTheta_g^{(k+1/2)}=\bI_q-\bbeta_g^{(k)}\bLambda_g^{(k)}+\bbeta_g^{(k)}\bS_g^{(k+1)}\bbeta_g^{(k)'}$ is a symmetric $q\times q$ matrix.

\subsubsection{CM-step}

At the CM-step on the second cycle of the $(k+1)$th iteration, we maximize $Q_2(\bvartheta_2)$ with respect to $\bvartheta_2$, fixing $\bvartheta_1=\bvartheta_1^{(k+1)}$.
The resulting updates for $\bvartheta_2$, when we impose the covariance constraints of \tablename~\ref{tab:Family of models} on the $\bLambda_g$ and $\bPsi_g$ matrices, can be derived from the expression for $Q_2(\bvartheta_2)$. Outline calculations required to compute the updates for all of the eight parsimonious covariance structures are given below; further details can be found in \citet[][Appendix~A]{McNi:Murp:Pars:2008}. More precisely, let us define
\begin{displaymath}	
\bS^{(k+1)}=\frac{1}{n}\sum_{g=1}^G n_g^{(k+1/2)}\bS_g^{(k+1)} \quad \text{and} \quad \bTheta^{(k+1/2)}=\bI_q-\bbeta^{(k)}\bLambda^{(k)}+\bbeta^{(k)}\bS^{(k+1)}\bbeta^{(k)'},
\end{displaymath}
with $\bbeta^{(k)}$ defined according to the imposed constraints.
Then, we obtain the following update equations for the eight different cases considered.
\begin{itemize}
	\item For model CCC, $\bLambda_g=\bLambda$ and $\bPsi_g=\bPsi=\psi\bI_p$, and the updates are
	\begin{equation*}\begin{split}
\bbeta^{(k)} &= \bLambda^{(k)'}(\bLambda^{(k)}\bLambda^{(k)'}+\psi^{(k)}\bI_p)^{-1},\quad
\bLambda^{(k+1)} = \bS^{(k+1)} \bbeta^{(k)} (\bTheta^{(k+1/2)})^{-1},\\
\psi^{(k+1)} &= \frac{1}{p}\tr(\bS^{(k+1)}-\bLambda^{(k+1)}\bbeta^{(k)}\bS^{(k+1)}).
\end{split}\end{equation*}
	\item For model CCU, $\bLambda_g=\bLambda$ and $\bPsi_g=\bPsi$, and the updates are
	\begin{equation*}\begin{split}
\bbeta^{(k)}     &= \bLambda^{(k)'}(\bLambda^{(k)}\bLambda^{(k)'}+\bPsi^{(k)})^{-1},\quad
\bLambda^{(k+1)} = \bS^{(k+1)} \bbeta^{(k)} (\bTheta^{(k+1/2)})^{-1},\\
\bPsi^{(k+1)}    &= \diag(\bS^{(k+1)}-\bLambda^{(k+1)}\bbeta^{(k)}\bS^{(k+1)}).
\end{split}\end{equation*}
	\item For model CUC, $\bLambda_g=\bLambda$ and $\bPsi_g=\psi_g\bI_p$, and the updates are
	\begin{equation*}\begin{split}
\bbeta_g^{(k)} &= \bLambda^{(k)'}(\bLambda^{(k)}\bLambda^{(k)'} + \psi_g^{(k)}\bI_p)^{-1},\\
\bLambda^{(k+1)} &= \left[\sum_{g=1}^G\frac{n_g^{(k+1/2)}}{\psi_g^{(k)}}\bS_g^{(k+1)}\bbeta_g^{(k)'}\right]
\left[\sum_{g=1}^G\frac{n_g^{(k+1/2)}}{\psi_g^{(k)}}\bTheta_g^{(k+1/2)}\right]^{-1},\\
\psi_g^{(k+1)} &=  \frac{1}{p}\tr(\bS_g^{(k+1)}-2\bLambda^{(k+1)}\bbeta_g^{(k)}\bS_g^{(k+1)}+\bLambda^{(k+1)}\bTheta_g^{(k+1/2)}\bLambda^{(k+1)'}).
\end{split}\end{equation*}
	\item For model CUU, $\bLambda_g=\bLambda$, and the updates are
	\begin{equation*}\begin{split}
\bbeta_g^{(k)} &= \bLambda^{(k)'}(\bLambda^{(k)}\bLambda^{(k)'}+\bPsi_g^{(k)})^{-1},\quad
\blambda_h^{(k+1)} = \br_h^{(k+1/2)}\left[\sum_{g=1}^G\frac{n_g^{(k+1/2)}}{\displaystyle\psi_{gh}^{(k)}}\bTheta_g^{(k+1/2)}\right]^{-1},\\
\bPsi_g^{(k+1)} &= \diag(\bS_g^{(k+1)}-2\bLambda^{(k+1)}\bbeta_g^{(k)}\bS_g^{(k+1)}+\bLambda^{(k+1)}\bTheta_g^{(k+1/2)}\bLambda^{(k+1)'}),
\end{split}\end{equation*}
where, for $h=1,\ldots,p$, $\blambda_h^{(k+1)}$ is the $p$th row of the matrix $\bLambda^{(k+1)}$, $\psi_{gh}^{(k)}$ denotes the $h$th element along the diagonal of $\bPsi_g^{(k)}$, and $\br_h^{(k+1/2)}$ represents the $h$th row of the matrix 
\begin{displaymath}
\displaystyle\sum_{g=1}^Gn_g^{(k+1/2)}(\bPsi_g^{(k)})^{-1}\bS_g^{(k+1)}\bbeta_g^{(k)'}.
\end{displaymath}
\item For model UCC, $\bPsi_g=\bPsi=\psi\bI_p$, and the updates are 
	\begin{equation*}\begin{split}
\bbeta_g^{(k)} &= \bLambda_g^{(k)'}(\bLambda^{(k)}_g\bLambda^{(k)'}_g+\psi^{(k)}\bI_p)^{-1},\quad
\bLambda_g^{(k+1)} = \bS_g^{(k+1)} \bbeta_g^{(k)'} (\bTheta_g^{(k+1/2)})^{-1},\\
\psi^{(k+1)} &= \frac{1}{np}\sum_{g=1}^Gn_g^{(k+1/2)}\tr(\bS_g^{(k+1)}-\bLambda_g^{(k+1)}\bbeta_g^{(k)}\bS_g^{(k+1)}).
\end{split}\end{equation*}
\item For model UCU, $\bPsi_g=\bPsi$, and the updates are 
	\begin{equation*}\begin{split}
\bbeta_g^{(k)} &= \bLambda_g^{(k)'}(\bLambda^{(k)}_g\bLambda^{(k)'}_g+\bPsi^{(k)})^{-1},\quad
\bLambda_g^{(k+1)} = \bS_g^{(k+1)} \bbeta_g^{(k)'} (\bTheta_g^{(k+1/2)})^{-1},\\
\bPsi^{(k+1)} &= \frac{1}{n}\sum_{g=1}^Gn_g^{(k+1/2)}\diag(\bS_g^{(k+1)}-\bLambda_g^{(k+1)}\bbeta_g^{(k)}\bS_g^{(k+1)}).
\end{split}\end{equation*}
\item For model UUC, $\bPsi_g=\psi_g\bI_p$, and the updates are 
	\begin{equation*}\begin{split}
\bbeta_g^{(k)} &= \bLambda_g^{(k)'}(\bLambda^{(k)}_g\bLambda^{(k)'}_g+\psi_g^{(k)}\bI_p)^{-1},\quad
\bLambda_g^{(k+1)} = \bS_g^{(k+1)} \bbeta_g^{(k)'} (\bTheta_g^{(k+1/2)})^{-1},\\
\psi_g^{(k+1)} &= \frac{1}{p}\tr(\bS_g^{(k+1)}-\bLambda_g^{(k+1)}\bbeta_g^{(k)}\bS_g^{(k+1)}).
\end{split}\end{equation*}
\item For model UUU, there are no constraints and the updates are
	\begin{equation*}\begin{split}
\bbeta_g^{(k)} &= \bLambda_g^{(k)'}(\bLambda^{(k)}_g\bLambda^{(k)'}_g+\bPsi_g^{(k)})^{-1},\quad
\bLambda_g^{(k+1)} = \bS_g^{(k+1)} \bbeta_g^{(k)'} (\bTheta_g^{(k+1/2)})^{-1},\\
\bPsi_g^{(k+1)} &= \diag(\bS_g^{(k+1)}-\bLambda_g^{(k+1)}\bbeta_g^{(k)}\bS_g^{(k+1)}).
\end{split}\end{equation*}
\end{itemize}

\section{Further computational details}
\label{sec:Computational details}

\subsection{Initialization}
\label{subsec:initialization}

The choice of the starting values for the AECM algorithm constitutes an important issue.
%
Instead of selecting $\bvartheta^{\left(0\right)}$ randomly, we suggest the following technique.
The mixture of Gaussian factor analyzers (MGFA) model, with a particular parsimonious covariance structure, can be seen as nested in four MCGFA models, those having the same parsimonious covariance structure.
In particular, the former can be obtained from the latter when $\alpha_g\rightarrow 1^-$ (or $\alpha\rightarrow 1^-$) and $\eta_g\rightarrow 1^+$ (or $\eta\rightarrow 1^+$), $g = 1, \ldots, G$.
Based on this idea, for all the four members of the MCGFA family having the same parsimonious covariance structure, the AECM algorithm is initialized with the estimates of $\{\pi_g,\bmu_g, \bLambda_g, \bPsi_g\}_{g=1}^G$ provided by the corresponding MGFA model, with same constraints set upon $\{\bSigma_g\}_{g=1}^G$.
The contamination parameters are initialized with fixed values close to, but not exactly 1, to avoid singularities in the first iteration of the algorithm. 
In our implementation we initialize with $\alpha_g^{(0)} = \alpha^{(0)} = 0.999$ and $\eta_g^{(0)} = \eta^{(0)} = 1.001$, $g = 1, \ldots, G$.
The (preliminary) MGFA model is estimated using the \texttt{pgmmEM()} function of the \textbf{pgmm} package for \textsf{R} \citep{McNi:Jamp:McDa:Murp:Bank:pgmm:2011}.
The \texttt{pgmmEM()} function implements an AECM algorithm to obtain ML estimates, and fitting models with the same covariance constraints as the MCGFA models.
In turn, to initialize this algorithm, we use an 
emEM strategy, for each $G$ and $q$, where 26 starts are run (25 random plus one $k$-means) for 5 iterations each using the unconstrained model 
and the start that led to the best BIC is pursued. 
See \cite{biernacki03} for further details on the emEM approach.
Initial parameter estimates are then computed componentwise via the EM updates.  

From an operational point of view, thanks to the monotonicity property of the AECM algorithm, this nested relation between MGFA and MCGFA models also guarantees that the observed-data log-likelihood of the MCGFA model will be always greater than, or equal to, the observed-data log-likelihood of the corresponding MGFA model.
This is a fundamental consideration for the use of likelihood-based criteria for selecting between these mixtures \citep{Punz:Brow:McNi:Hypo:2016}.


\subsection{Convergence Criterion}
\label{subsubsec: CN: Convergence criterion}

The Aitken acceleration \citep{aitken26} is used to estimate the asymptotic maximum of the log-likelihood at each iteration of the AECM algorithm. 
Based on this estimate, we can decide whether or not the algorithm has reached convergence; i.e., whether or not the log-likelihood is sufficiently close to its estimated asymptotic value. 
The Aitken acceleration at iteration $k+1$ is given by
\begin{equation*}
a^{\left(k+1\right)}=\frac{l^{\left(k+2\right)}-l^{\left(k+1\right)}}{l^{\left(k+1\right)}-l^{\left(k\right)}},
\end{equation*}
where $l^{\left(k\right)}$ is the observed-data log-likelihood value from iteration $k$.  
Then, the asymptotic estimate of the log-likelihood at iteration $k + 2$ is given by
\begin{displaymath}	
l_{\infty}^{\left(k+2\right)}=l^{\left(k+1\right)}+\frac{1}{1-a^{\left(k+1\right)}}(l^{\left(k+2\right)}-l^{\left(k+1\right)});
\end{displaymath}
see \citet{Bohn:Diet:Scha:Schl:Lind:TheD:1994}.
The AECM algorithm can be considered to have converged when $l_{\infty}^{\left(k+2\right)}-l_{\infty}^{\left(k+1\right)}<\epsilon$, where $\epsilon$ is the desired tolerance. 

\subsection{Woodbury identity}
\label{subsec: MCGFAM: Computational details}

The second cycle E-step of the AECM algorithm, in the computation of $\bbeta_g^{\left(k\right)}$, requires the inversion of the $p\times p$ matrix $\bLambda_g^{\left(k\right)}\bLambda_g^{\left(k\right)'}+\bPsi_g^{\left(k\right)}$, $g=1,\ldots,G$.
This inversion can be slow for large values of $p$.
To ease it, we use the Woodbury identity \citep{Wood:Inve:1950}
\begin{equation}
\left(\bLambda_g^{\left(k\right)}\bLambda_g^{\left(k\right)'}+\bPsi_g^{\left(k\right)}\right)^{-1}
=
\left(\bPsi_g^{\left(k\right)}\right)^{-1} - \left(\bPsi_g^{\left(k\right)}\right)^{-1}\bLambda_g^{\left(k\right)}\left[\bI_q+\bLambda_g^{\left(k\right)'}\left(\bPsi_g^{\left(k\right)}\right)^{-1}\bLambda_g^{\left(k\right)}\right]^{-1}\bLambda_g^{\left(k\right)'}\left(\bPsi_g^{\left(k\right)}\right)^{-1},
\label{eq:Woodbury identity}
\end{equation}
which requires the simpler inversions of the diagonal $p\times p$ matrix $\bPsi_g^{\left(k\right)}$ and the $q\times q$ matrix $\bI_q+\bLambda_g^{\left(k\right)'}\left(\bPsi_g^{\left(k\right)}\right)^{-1}\bLambda_g^{\left(k\right)}$. 
This leads to a particularly significant speed-up when $q \ll p$.




%

\section{Comparison with competing methods}
\label{sec:Comparison with competing methods}

In this section, we compare the clustering and classification performance of the MCGFA model to two natural competitors. 
\begin{description}
	\item[EPGMM] is the expanded parsimonious Gaussian mixture model family (EPGMM), introduced by \citet{mcnicholas:murphy:2010:epgmm}. 
EPGMM is a 12-member family of MGFA models, that extends the 8-member PGMM family of \citet{McNi:Murp:Pars:2008}. 
Model fitting for EPGMM was implemented by the \texttt{pgmmEM()} function of the \textbf{pgmm} package.
	\item[MMtFA] is the family of mixtures of modified $t$-factor analyzers (MMtFA) models of \citet{andrews11c}. 
MMtFA is a 24-member family of mixtures of factor analyzers models based on the multivariate $t$-distribution as opposed to the Gaussian. 
The 24 models are analogous to the 12 models of the EPGMM family, with an additional possible constraint on the degrees of freedom parameter doubling the number of possibilities. 
Model fitting for MMtFA was implemented by the \texttt{mmtfa()} function of the \textbf{mmtfa} package for \R \citep{Andrews:McNicholas:Chalifour:mmtfa:2015}.
\end{description}
Mixtures of modified $t$-factor analyzers are the closest competitor to MCGFA; both models are factor analysis models based off of heavy-tailed elliptical distributions. 
The inherent advantage of the MCGFA model is that bad points are, if required, automatically and explicitly identified. 
The MMtFA model instead assimilates bad points into clusters.
An additional advantage of the MCGFA is a simplified AECM algorithm. 
Numerical optimization is necessary in the equivalent algorithm for MMtFA model because there is no closed-form update available for the degrees-of-freedom parameter in each cluster. 
The MCGFA model was applied using the emEM initialization strategy described in Section~\ref{subsec:initialization}.

For completeness, it is worth noting that trimming approaches based on Gaussian factor analyzers have been developed for use in high-dimensional clustering problems \citep[see][]{garcia2016joint,Yang:Xian:Yao:Robu:2017}. 
While these approaches can be effective if provided with the correct proportion of outlying points, the need to pre-specify the proportion of outlying points greatly limits the extent to which they can be used in comparisons. 
Specifically, while it is straightforward to make a good guess at the proportion of outlying points in very low dimensions (e.g., $p=2$ or $p=3$), there is no reliable way to do so in general. 
Therefore, we limited our comparisons to relevant approaches that do not require pre-specification of the proportion of bad points.
  
For each application, every member of each family of models \textcolor{blue}{was fitted} with a range of values for $G$ and $q$, and the best model for family \textcolor{blue}{was selected} using BIC (cf.~Section~\ref{subsec:Model selection}). 
Thus each application of the \mcgfa, \mmtfa and \pgmm ``methods'' \textcolor{blue}{involved} many models with different covariance structures, numbers of components and numbers of latent factors and choosing the best one.
Thus the methods can be evaluated on both model fitting and the success of the BIC model selection procedure.

To be precise, the methods \textcolor{blue}{were judged} on their ability to:
\begin{enumerate}[label=\roman*.]
\item separate known clusters;
\item recover known structure in the data ($G$ and $q$) through model selection;
\item produce parsimonious models with the best overall fit to the data.
\end{enumerate}
The first criterion \textcolor{blue}{was measured} using the adjusted Rand index \citep[ARI;][]{hubert85}, which is a measure of agreement between partitions that is applicable even to partitions of differing sizes. 
An ARI value of 1 indicates perfect agreement, and the expected value of the ARI under random classification is 0. 
When the methods \textcolor{blue}{were applied} to data with known labels, the results \textcolor{blue}{were evaluated} against this reference. 
The second point is straightforward: when the true values of $G$ or $q$ \textcolor{blue}{were known}, we \textcolor{blue}{saw} whether they \textcolor{blue}{matched} the corresponding values in the selected models. 
The third criterion \textcolor{blue}{was measured} by comparing the BIC value directly. 
The BIC rewards models that closely fit the data, but penalizes models that are highly parameterized and may suffer from overfitting (cf.~Section~\ref{subsec:Model selection}).
It is worth noting that the MCGFA family of models is inherently less parsimonious than the MMtFA family because
the contaminated Gaussian distribution has an additional parameter compared to the $t$-distribution.
Thus the BIC values for the MCGFA may \textcolor{blue}{tend to be higher} than those of the MMtFA. 
On the other hand, the MCGFA model uses these parameters to provide automatic classification of bad points. 
Therefore, in addition to the above criteria, the MCGFA method \textcolor{blue}{was evaluated} on its ability to detect such points, when appropriate.

In every case, the data \textcolor{blue}{were scaled} to have mean 0 and standard deviation 1 on each variate before the fitting methods \textcolor{blue}{were applied}.
This is the approach recommended by the \textbf{mmtfa} package. 
Scaling is generally considered good practice, does not change cluster shape, removes the impact of measurement unit, and also \textcolor{blue}{helps} avoid numerical issues affecting the convergence of the fitting algorithm.

\subsection{Simulated data analysis}

In this section, five types of simulated data sets \textcolor{blue}{were considered}:
\begin{enumerate}
	\item Gaussian clusters;
	\item Contaminated Gaussian clusters;
	\item $t$-distributed clusters;
	\item Gaussian clusters with noise;
	\item Example with $p=100$.
\end{enumerate}
In all cases, there were $G=2$ components and each component had a latent factor structure with $q=3$ latent factors. 
In the first four cases, ten replications of $p=10$ dimensional data with equally sized components ($\pi_1=\pi_2=0.5$) were generated with $n=200$ as sample size. 
Other settings varied per example and details are provided in the relevant section. 
In each case, every parsimonious model in each of the MCGFA, MMtFA and EPGMM families was fitted with $G =1,\ldots,5$ components and $q = 1,\ldots,5$ latent factors, and the best model in each family was selected by the BIC.

\subsubsection{Gaussian clusters}
\label{sec:thisone}

The first component had mean at the origin and the second had a mean vector drawn from a Gaussian distribution centered at the origin with covariance matrix $\I_{10}$.
Two loading matrices, $\bLambda_1$ and $\bLambda_2$, were generated with components drawn from independent Gaussian distributions centered at the origin with covariance matrix $\I_{10}$.
The elements on the diagonals of the error variance matrices, $\bPsi_1$ and $\bPsi_2$, were randomly generated from a uniform distribution on $(0.5,1)$. 
All three approaches (\mcgfa, \mmtfa and \pgmm) were run on all ten datasets.
The BIC selected $G = 2$ components for all models on all runs and the classification performance for all methods was very good, with the \mcgfa approach having a slightly higher mean ARI (\tablename~\ref{tab:gauss:cls}). 
For all MCGFA and MMtFA models, $q=3$ latent factors were selected for all runs and, for the EPGMM family, $q=3$ latent factors were selected on 9 of the 10 runs (\tablename~\ref{tab:gauss:q}). 
\begin{table*}[!ht]
	\centering
	\caption{Mean ARI and BIC values, with respective standard deviations in parentheses, for the mixtures of factor analyzers models on Gaussian clusters (std.~errors in parentheses).}
	\label{tab:gauss:cls}
		\begin{tabular}{lccc}
			\toprule
			& \mcgfa & \mmtfa & \pgmm \\   
			\midrule
			Mean ARI    & \textbf{0.872} (0.04)  & 0.867 (0.05)   &  0.863  (0.04)  \\
			Mean BIC    & 4711.48 (66.1) & \textbf{4698.43} (60.5) & 4726.32 (68.4)   \\
			\bottomrule
		\end{tabular}
\end{table*}
\begin{table}[!ht]
	\centering
	\caption{Number of latent factors $q$ selected by BIC on Gaussian clusters.}
	\label{tab:gauss:q}
		\begin{tabular}{lcccc}
			\toprule
			$q$    & \mcgfa & \mmtfa & \pgmm \\   
			\midrule
			1      &  0    &    0   &  0  \\
			2      &  0    &   0    &  0  \\
			\rowcolor{lightgray}
			3      & 10     & 10     & 9    \\
			4      &  0    &  0    & 1    \\
			5      &  0    &   0    &  0  \\
			\bottomrule
		\end{tabular}
\end{table}

\subsubsection{Contaminated Gaussian clusters}
\label{sec:thisone2}

The data were generated in the same way as in Section~\ref{sec:thisone} but with a covariance inflation factor $\eta_g$ for each component drawn from an exponential distribution (truncated at 1) with mean 10. 
Ten percent of observations in the first group and twenty percent of those in the second group were designated as ``bad'', i.e., $\alpha_1 =0.9$ and $\alpha_2 = 0.8$.
Each combination of these randomly generated parameters yielded a pair of contaminated Gaussian clusters. 
All three approaches (\mcgfa, \mmtfa and \pgmm) were run on all ten datasets.
Unsurprisingly, the \mcgfa approach gave the best performance in terms of both mean BIC and mean ARI (\tablename~\ref{tab:cg:cls}). 
The classification performance of the \mmtfa approach was similar but the EPGMM approach did not perform as well, which illustrated the deleterious impact of the outliers (\tablename~\ref{tab:cg:cls}).
In all cases, the \mcgfa and \mmtfa approaches selected a model with $G=2$ components but \pgmm needed additional components to help account for the outliers (\tablename~\ref{tab:cg:mod}). 
This time, the number of latent factors selected in each case was the same as for the previous simulation (see \tablename~\ref{tab:gauss:q}) and so is not repeated here.
\begin{table*}[!ht]
	\centering
	\caption{Mean ARI and BIC values, with respective standard deviations in parentheses, for the mixtures of factor analyzers models on contaminated Gaussian clusters.}
	\label{tab:cg:cls}
		\begin{tabular}{lccc}
			\toprule
			& \mcgfa & \mmtfa & \pgmm \\   
			\midrule
			Mean ARI & \textbf{0.957} (0.03) & 0.945 (0.03) & 0.756 (0.08) \\
			Mean BIC & \textbf{3175.11} (311) & 3259.95 (305) & 3480.40 (322) \\
			\bottomrule
		\end{tabular}
\end{table*}
\begin{table}[!ht]
	\centering
	\caption{Model selection performance of BIC on contaminated Gaussian clusters.}
	\label{tab:cg:mod}
		\begin{tabular}{lcccc}
			\toprule
			$G$    & \mcgfa & \mmtfa & \pgmm \\   
			\midrule
			1      & 0       & 0      & 0     \\
			\rowcolor{lightgray}
			2      & \textbf{10}        & \textbf{10}     & 0    \\
			3      & 0         &0      & 8     \\
			4      & 0      &0    &  2 \\
			5      & 0      &  0&  0 \\
			\bottomrule
		\end{tabular}
\end{table}

\subsubsection{$t$-distributed clusters}

The means and scale matrices were generated as in Section~\ref{sec:thisone}. 
The degrees of freedom parameters were set to $\nu_1=10$ and $\nu_2=60$, respectively. 
In all cases, the BIC selected $G=2$ components, classification performance was very good (\tablename~\ref{tab:t:cls0}), and the number of factors was usually $q=3$ (\tablename~\ref{tab:t:q}). 
\begin{table*}[!ht]
	\centering
	\caption{Clustering performance of factor analyzer models on $t$-distributed clusters.}
	\label{tab:t:cls0}
	\begin{tabular}{lccc}
		\toprule
		& \mcgfa & \mmtfa & \pgmm \\   
		\midrule
		Mean ARI & 0.904 (0.03) & 0.908 (0.03) & \textbf{0.914} (0.02) \\
		Mean BIC & 4577.27 (94.4) & \textbf{4566.86} (89.1) & 4593.01 (87.1)  \\
		\bottomrule
	\end{tabular}
\end{table*}
\begin{table}[!ht]
	\centering
	\caption{Number of latent factors $q$ selected by BIC on contaminated Gaussian clusters.}
	\label{tab:t:q}
		\begin{tabular}{lccc}
	\toprule
	$q$    & \mcgfa & \mmtfa & \pgmm \\   
	\midrule
	1 & 0       & 0     &   0    \\
	2      & 0       & 0     &   1    \\
	\rowcolor{lightgray}
	3      & 9      & \textbf{10}    & 9    \\
	4      & 1       & 0     & 0     \\
	5      & 0       & 0     & 0 \\
	\bottomrule
\end{tabular}
\end{table}

Unsurprisingly, the MMtFA model had the best mean BIC. 
However, the EPGMM had a slightly higher average ARI value. 
This is not surprising when one considers that the second component was effectively Gaussian (i.e., $\nu_2=60$) while the first was not particularly heavy tailed (i.e., $\nu_1=10$).

\subsubsection{Gaussian clusters with uniform noise}
\label{subsec:Gaussian clusters with uniform noise}

The means and covariance matrices were generated as in Section~\ref{sec:thisone}. 
Then, 20 noise points were added to the data, drawn uniformly from $(-5,5) \times \cdots \times (-5,5)$. 
The noise observations were not considered in the evaluation of clustering performance. 
The BIC selected $G=2$ components for all models.
All three approaches gave very good classification performance and the MCGFA approach was the best, albeit by a small margin (\tablename~\ref{tab:noise:cls}). 
The MMtFA models were the only ones that consistently had $q=3$ latent factors (\tablename~\ref{tab:noise:q}). 
\begin{table*}[!ht]
	\centering
    \caption{Clustering performance of factor analyzer models on Gaussian clusters with uniform noise.}
    \label{tab:noise:cls}
	\begin{tabular}{lccc}
		\toprule
		& \mcgfa & \mmtfa & \pgmm \\   
		\midrule
		Mean ARI & \textbf{0.936} (0.06) & 0.926 (0.05) & 0.902 (0.05) \\
		Mean BIC & \textbf{5342.07} (54.28) & 5346.38 (43.86) & 5464.56 (73.78)  \\
		\bottomrule
	\end{tabular}
\end{table*}
\begin{table}[!ht]
	\centering
	\caption{Number of latent factors ($q$) selected by BIC on Gaussian clusters with uniform noise.}
	\label{tab:noise:q}
		\begin{tabular}{lccc}
			\toprule
			$q$    & \mcgfa & \mmtfa & \pgmm \\   
			\midrule
			1 & 0       & 0     & 0 \\
			2      & 0       & 0     &   3    \\
			\rowcolor{lightgray}
			3      & 6      & \textbf{10}    & 7    \\
			4      & 4       & 0     & 1     \\
			5      & 0       & 0     & 0 \\
			\bottomrule
		\end{tabular}
\end{table}

In addition to clustering performance, the \mcgfa model was judged on its ability to detect ``bad'' points. 
Both \textit{sensitivity} and \textit{specificity} were considered. 
The sensitivity is the proportion of bad points successfully detected, and the specificity is the proportion of good points successfully labelled as such. 
The detection results for each initialization scheme of our models are shown in \tablename~\ref{tab:detect1}. 
The specificity figures were impressive considering noise points might easily lie within clusters.
\begin{table}[!ht]
	\centering
	\caption{Outlier detection results for the MCGFA method on Gaussian clusters with uniform noise.}
	\label{tab:detect1}
		\begin{tabular}{lr}
			\toprule
			Mean \# Correctly Detected & 19.3 \\
			Mean \# Falsely Detected & 6.9  \\
			Mean Sensitivity & 96.5\%  \\
			Mean Specificity & 96.6\% \\
			\bottomrule
		\end{tabular}
\end{table}

\subsubsection{Example with $p=100$}

Ten replications of $p=100$ dimensional data with $G=2$ equally sized ($\pi_1=\pi_2=0.5$) components were generated in each case.
A total of $n = 440$ observations were generated in each case: 400 regular plus 40 noise. 
In each dimension, the noise was uniform on $(-15,15)$.
We set $\bmu_1$ at the origin, and $\bmu_2$ at the origin in each dimension but the first 10 dimensions, where it took a value 4.
Note that, here, the true model was a CUU model from the PGMM family.
Elements of the factor loading matrix $\bLambda$ were generated randomly from a standard Gaussian distribution. 
The diagonal elements of $\bPsi_1$ and $\bPsi_2$ were randomly generated from a uniform on $(0.5, 10)$.
This time, each model was fitted for $G = 1,\ldots,5$ and $q = 1,\ldots,10$.
Each model chose $G = 3$ and $q = 5$ for all 10 simulations, and gave perfect clustering results (with noise appearing as a separate cluster for both MMtFA and PGMM).
In all cases, the chosen MCGFA model is CUUUC, the MMTFA model is CUUC, and the PGMM is CUU --- these all make sense considering that the data are generated form a PGMM CUU model.

While this turned out to be a relatively straightforward clustering problem, there are three interesting takeaways: 
the MCGFA model work well in high dimensions; 
by BIC, MCGFA outperformed mmtfa (\tablename~\ref{tab:100:cls}); 
even with extra parameters, MCGFA outperformed PGMM by BIC in half the replications.
\begin{table*}[!ht]
	\centering
    \caption{Average BIC, and number of times the BIC value is the smallest (No.\ min.\ BIC), for mixtures of factor analyzers models for the $p=100$ example.}
    \label{tab:100:cls}
	\begin{tabular}{lccc}
		\toprule
		& \mcgfa & \mmtfa & \pgmm \\   
		\midrule
		Mean BIC & 90472.22 & 90515.55 & \textbf{90468.04}\\
		No.\ min.\ BIC & 5 & 5 & 0\\
		\bottomrule
	\end{tabular}
\end{table*}

\subsection{Real data analyses}

\subsubsection{Wine data set}

The wine data set \citep{forina86} consists of $p=27$ chemical properties of $n=178$ bottles of wine, of three different types: Barolo, Grigolino and Barbera. 
The data set is available in the \textbf{pgmm} package for \R. 
Each method was fitted to the data with every set of constraints, $G = 1, \ldots, 10$ components and $q = 1,\ldots,10$ latent factors. 
The results (\tablename s~\ref{tab:winetables} and \ref{tab:wineperf}) show that all three approaches gave very good classification performance with the MCGFA and MMtFA models slightly outperforming the EPGMM model. 
Interestingly, the scale matrices had the same (CUU) structure in each case.
\begin{table*}[!ht]
	\centering
	\caption{Contingency tables for mixtures of factor analyzers models applied to the wine data.}
	\label{tab:winetables}
		\begin{tabular}{lccccccccccc}
			\toprule
			&  \multicolumn{3}{c}{\mcgfa} && \multicolumn{3}{c}{\mmtfa} && \multicolumn{3}{c}{\pgmm}  \\
			\cline{2-4}\cline{6-8}\cline{10-12}
			& 1  & 2  & 3  && 1  & 2  & 3  && 1  & 2  & 3 \\
			\midrule
			Barolo		& 59 & 0  & 0 & & 59 & 0  & 0  && 59 & 0  & 0  \\
			Grignolino	& 2  & 69 & 0  && 2  & 69 & 0  && 3  & 67 & 1 \\
			Barbera		& 0  & 0  & 48 && 0  & 0  & 48 && 0  & 0  & 48  \\
			\bottomrule
		\end{tabular}
\end{table*}
\begin{table}[!ht]
	\centering
	\caption{Details of the chosen model for each mixtures of factor analyzers approach applied to the wine data.}
	\label{tab:wineperf}
		\begin{tabular}{lccc}
			\toprule
			& \mcgfa	& \mmtfa	& \pgmm	\\   
			\midrule
			model 	& CUUCC	& CUUC		& CUU	\\
			$q$ 		& 4				& 4			& 6		\\
			ARI		& \textbf{0.964} 		& \textbf{0.964}		& 0.929   \\
			BIC		& 11347.82 &    \textbf{11339.23}  & 11479.09  \\
			\bottomrule
		\end{tabular}
\end{table}

The \mmtfa model achieved the best BIC value. 
This is \textcolor{blue}{probably} because the larger \mmtfa family includes some parsimonious models that have no analogue in the \mcgfa family. 
The best \mmtfa model was CUUC; the final ``C'' indicates that the degrees of freedom parameter was held equal across the groups so there is only one parameter in the model that controls the shape of the tails of the component distributions. 
Meanwhile, the best \mcgfa model had 6 parameters ($\alpha_g$ and $\eta_g$, $g=1,2,3$) for the same task.

To explore the effect of outliers on model performance, a new version of the wine data was created by adding two artificial observations. 
These observations were generated by copying the first two observations from the Barolo group and giving them an alcohol level of 25\%. 
The results highlighted an interesting advantage of the MCGFA approach in this situation (\tablename s~\ref{tab:winetables2} and \ref{tab:wineperf2}). 
As one would expect, the EPGMM approach did not perform well. 
The MCGFA and MMtFA might seem to give similar performance but it is important to note that only the MCGFA selected the correct (i.e., same as before) covariance structure despite the outliers. 
\begin{table*}[!ht]
	\centering
	\caption{Contingency tables for mixtures of factor analyzers models applied to the contaminated wine data.}
	\label{tab:winetables2}
		\begin{tabular}{lcccccccccccccccccc}
			\toprule
			&  \multicolumn{3}{c}{\mcgfa} && \multicolumn{3}{c}{\mmtfa} && \multicolumn{10}{c}{\pgmm}  \\
			\cline{2-4}\cline{6-8}\cline{10-19}
			& 1  & 2  & 3  && 1  & 2  & 3  && 1  & 2  & 3 &4 & 5 & 6 & 7 & 8 & 9 & 10 \\
			\midrule
			Barolo		& 59 & 0  & 0 & & 59 & 0  & 0  && 15 & 12 & 22 &  7 &  1 &  4 &  0 &  0 &  0 &  0 \\
			Grignolino		& 2  & 69 & 0  && 3  & 68 & 0  &&  0 &  0 & 15 &  3 & 14 &  0 & 34 &  5 &  0 &  0 \\
			Barbera		& 0  & 0  & 48 && 0  & 0  & 48 &&  0 &  0 &  0 &  0 &  0 &  0 &  0 &  4 & 39 &  5 \\
			\bottomrule
		\end{tabular}
\end{table*}
\begin{table}[!ht]
	\centering
	\caption{Performance measures for each mixtures of factor analyzers models applied to the contaminated wine data.}
	\label{tab:wineperf2}
		\begin{tabular}{lccc}
			\toprule
			& \mcgfa	& \mmtfa	& \pgmm	\\   
			\midrule
			model 	& CUUUU	& UCUC		& UUU	\\
			$q$ 		& 4				& 4			& 1		\\
			ARI		& \textbf{0.964} 		& 0.946		& 0.376   \\
			BIC		& 11405.55 &    \textbf{11392.38}  & 11392.42  \\
			\bottomrule
		\end{tabular}
\end{table}

\subsubsection{AIS data set}

The Australian Institute of Sport data set \citep{Cook:Weisberg:1994} consists of $p=11$ numerical measurements of $n=202$ athletes, along with their classification by gender and sport. 
There are 9 women's sports and 8 men's sports, for a total of 17 nested classes. 
The ratio of observations to classes is too low to hope to uncover the 17 component structure, so we evaluated the models primarily based on their ability to separate the athletes by gender. 
However, we also investigated how each method partitions athletes with regards to sport.
Each method was fitted to the data with every set of constraints, $G = 1, \ldots, 10$ components and $q = 1,\ldots,5$ latent factors. 
The classification results (\tablename s~\ref{tab:aistables} and~\ref{tab:aisperf}) show that all approaches selected a $G=3$ component model. 
The relatively poor classification performance of all approaches was unsurprising when one considers that the clusters in these data are well known to be asymmetric \citep[see, e.g.,][Chp.~7]{mcnicholas16a}. 
\begin{table}[!ht]
	\centering
	\caption{Contingency tables for mixtures of factor analyzers models applied to the AIS data set, by gender.}
	\label{tab:aistables}
		\begin{tabular}{lccccccccccc}
			\toprule
			&  \multicolumn{3}{c}{\mcgfa} && \multicolumn{3}{c}{\mmtfa} && \multicolumn{3}{c}{\pgmm}  \\
			\cline{2-4}\cline{6-8}\cline{10-12}
			& 1  & 2  & 3  && 1  & 2  & 3  && 1  & 2  & 3 \\
			\midrule
			female & 64 & 36 &  0 && 65 & 35 &  0 && 80 & 20 & 0  \\ 
			male   &  3 & 15 & 84 &&  3 & 16 & 83 &&  1 & 17 & 84 \\ 
			\bottomrule
		\end{tabular}
\end{table} 
%
\begin{table}[!ht]
	\centering
	\caption{Performance measures for mixtures of factor analyzers models applied to the AIS data set.}
	\label{tab:aisperf}
		\begin{tabular}{lcccc}
			\toprule
			& \mcgfa & \mmtfa & \pgmm \\   
			\midrule
			Model & UCUCU & UCCC & UUU \\
			$q$ & 4 & 5 & 4 \\
			ARI & 0.55 & 0.54 & \textbf{0.65}  \\
			BIC & \textbf{2219.921} & 2254.758 & 2312.379  \\
			\bottomrule
		\end{tabular}
\end{table}

To further investigate these results, we examined the contingency table of each clustering by each athlete's gender and sport (\figurename~\ref{tab:aistablesSport}). 
For both the MCGFA and MMtFA models, the second cluster contained a similar mix of genders (36/15 and 35/16, respectively) and so it was interesting to briefly consider these clusters. 
In both cases, the second cluster contained: female athletes who did neither field nor gymnastics plus male athletes who generally (but not exclusively) did swimming or 400m running.
\begin{table*}[!ht]
	\centering
	\caption{Contingency tables for mixtures of factor analyzers models applied to the AIS data set, by gender and sport}
	\label{tab:aistablesSport}
		\begin{tabular}{llccccccccccc}
			\toprule
			&&  \multicolumn{3}{c}{\mcgfa} && \multicolumn{3}{c}{\mmtfa} && \multicolumn{3}{c}{\pgmm}  \\
			\cline{3-5}\cline{7-9}\cline{11-13}
			& 				& 1 & 2 & 3 && 1 & 2 & 3 && 1 & 2 & 3 \\ 
			\midrule
  Female	& Row     &  12 & 10 & 0 && 13 & 9 & 0 && 15 & 7 & 0\\
            & Netball &  21 & 2 & 0 && 21 & 2 & 0 && 19 & 4 & 0\\  
            & BBall   &  12 & 1 & 0 && 12 & 1 & 0 && 11 & 2 & 0\\ 
            & Field   &  7 & 0 & 0 && 7 & 0 & 0 && 7 & 0 & 0\\ 
            & Swim    &  6 & 3 & 0 && 6 & 3 & 0 && 4 & 5 & 0\\  
            & Tennis  &  2 & 5 & 0 && 2 & 5 & 0 && 5 & 2 & 0\\
            & Gym     &  4 & 0 & 0 && 4 & 0 & 0 && 4 & 0 & 0\\  
            & TSprnt  &  0 & 4 & 0 && 0 & 4 & 0 && 4 & 0 & 0\\ 
            & T400m   &  0 & 11 & 0 && 0 & 11 & 0 && 11 & 0 & 0\\
\midrule
Male	    & Row    & 0 & 0 & 15 && 0 & 0 & 15 && 0 & 1 & 14 \\
            & WPolo  & 0 & 0 & 17 && 0 & 0 & 17 && 0 & 4 & 13 \\
            & BBall  & 0 & 1 & 11 && 0 & 1 & 11 && 0 & 1 & 11 \\
            & Field  & 0 & 1 & 11 && 0 & 1 & 11 && 1 & 1 & 10 \\
            & Swim   & 0 & 5 & 8 && 0 & 5 & 8 && 0 & 2 & 11\\
            & TSprnt & 0 & 0 & 4 && 0 & 0 & 4 && 0 & 0 & 4\\
            & Tennis & 2 & 1 & 8 && 2 & 1 & 8 && 0 & 5 & 6\\
            & T400m  & 1 & 7 & 10 && 1 & 8 & 9 && 0 & 3 & 15\\
			\bottomrule
		\end{tabular}
\end{table*}

\section{Discussion}
\label{sec:Discussion and future work}

In this paper, methodological contributions have been contextualized in the high-dimensional setting and have mainly involved the definition of both the contaminated Gaussian factor analysis (CGFA) model --- as a generalization of the classical (Gaussian) factor analysis model --- and the mixture of contaminated Gaussian factor analyzers (MCGFA) model.
In the fashion of \citet{McNi:Murp:Pars:2008} and \citet{R:teigen:JSS}, a family of 32 parsimonious MCGFA models has been also introduced that allow different constraints to be placed on to the factor loading, error variance matrices, and contamination parameters of different components in the mixture. 
These parsimonious variants 
provide smaller, more easily interpretable models.
In one sense, the CGFA model can be viewed as a generalization of the (Gaussian) factor analysis model, while the MCGFA model is a generalization of the mixture of (Gaussian) factor analyzers model. 
These generalizations aim to accommodate mild outliers which we have collectively referred to as bad points.
Although approaches for high-dimensional data, such as the $t$-factor analysis model and the mixture of $t$-factor analyzers model, can be used for data comprising bad points, they do not give the opportunity to automatically detect them.

Computational contributions have concerned the detailed illustration of AECM algorithms for fitting the above family of parsimonious MCGFA models.
A further advantage of the proposed approach over the mixture of $t$-factor analyzers model, in computational terms, is related to the fact that all of the parameters of the MCGFA model are available in a closed form in the iterations of the AECM algorithm, while the same does not hold for the mixture of $t$-factor analyzers model.
This avoids the use of numerical optimization for model fitting. 
Our MGCFA approach was compared to both the MMtFA and EPGMM approaches using real and simulation data. 
In each case, it gave either comparable or superior performance. 
While comparison to the MMtFA approach is interesting, it must be remembered that even when the performance is comparable, the MCGFA method yields automatic and explicit detection of bad points. 

There are several avenues for future work.
The models in our family assume that the bad (Gaussian) density in each cluster has the same shape of the good (Gaussian) density, but with an inflated covariance matrix.
While this results in a parsimonious model, some applications could require a more complex paradigm where good and bad densities have still the same mode, but are allowed to have a different shape.
If each mixture component is associated with a cluster, then the models in our family imply elliptically symmetric clusters, which may be rather restrictive in some real data applications. 
To overcome this problem, still preserving the possibility to reduce the dimensionality and to detect mild outliers, our 32 parsimonious configurations may be easily applied to the component scale matrices of mixtures of contaminated skewed distributions, such as mixtures of multivariate skew-contaminated normal distributions \citep{Cabr:Lach:Prat:Mult:2012} and mixtures of contaminated shifted asymmetric Laplace distributions \citep{Morr:Punz:McNi:Brow:Asym:2019}; for the use of skewed component distributions, see also, e.g., \citet{franczak14} and \citet{Punz:Mazz:Maru:Fitt:2018}. 
An analogous approach could be taken in other cases, such as the hypercube approach of \cite{franczak15}. 
Furthermore, analogous approaches to those we have used to develop the MCGFA family could be taken in the matrix variate case (see \citealp{viroli11} and \citealp{gallaugher18a}). 
Finally, ideas borrowed from high-dimensional work in the document domain (e.g., \citealp{Mark:Mill:Join:2010} and \citealp{soleimani2016atd}) could be applied to the mixture of factor analyzers model to produce an alternative approach --- it would be interesting to compare such an approach to the MCGFAs.

{\small\section*{Acknowledgements}
This work was  supported by the Canada Research Chairs program and an E.W.R. Steacie Memorial Fellowship (McNicholas).


}

\end{document}